\documentclass[traditabstract]{aa}
\usepackage{graphicx}
\usepackage{txfonts}
\usepackage{natbib}
\begin{document}

   \title{Stellar wobble caused by a binary system: \\
    Can it really be mistaken as an extra-solar planet?}

   \titlerunning{Stellar wobble caused by a binary system}
   \author{M.H.M. Morais
          \inst{1}
          \and
          A.C.M. Correia\inst{2,3}}

  \offprints{A.C.M. Correia, \email{correia@ua.pt}}

   \institute{Centro de F\'isica Computacional, Universidade de Coimbra,
              3004-516 Coimbra, Portugal \\
              \email{hmorais@mat.uc.pt}
         \and
             Departamento de F\'isica, Universidade de Aveiro,
	     Campus de Santiago, 3810-193 Aveiro, Portugal
         \and
             IMCCE, CNRS-UMR8028, Observatoire de Paris, UPMC, 77 avenue
             Denfert-Rochereau, 75014 Paris, France 
             }

   \date{Received \today; accepted \today}

 
  \abstract{The traditional method for detecting extra-solar planets relies on
  measuring a small stellar wobble which is assumed to be caused by a planet
  orbiting the star. Recently,  it has been suggested that a similar stellar wobble
  could be caused by a close binary system. Here we
  show that, although the effect of a close binary system can at first sight be
  mistaken as a planetary companion to the star, more careful analysis of the
  observational data should allow us to distinguish between the two effects.}

   \keywords{stars: planetary systems --
             stars: binaries: visual --
	     techniques: radial velocities -- 
             methods: observational  --   
	     celestial mechanics}

\maketitle
%

\section{Introduction}

Over the past decade up to 290 extra-solar planets were discovered using the wobble method
(i.e. measuring the motion of parent star around the planet-star's center of mass).
This represents about 95\% of the total extra-solar planets known to date\footnote{The Extrasolar Planets
Encyclopedia. http://exoplanet.eu/} 
However, this is an indirect method in which the presence of a planet is inferred from the detection of the radial velocity
variations. Therefore, in cases where the presence of the planet can not be confirmed by other methods we may ask if
these variations could be due to another dynamical effect.
In particular, when a planet is detected in a system already hosting a companion (such as another planet or a star), we must be 
cautious when we analyze the data, because different configurations of bodies can lead to similar radial velocity variations.

The present paper is motivated by the work of \citet{Schneider_Cabrera_2006} 
who studied a triple system composed of a binary
star system with equal masses orbiting a third star (assumed massless). They concluded that 
the binary system will cause a wobble in the star's motion
that could mimic the presence of a planet companion. Therefore,
they suggest that each extra-solar planet detection which relies on the
measurement of this wobble must be carefully checked for the presence of 
possible nearby unresolved binaries.

Unfortunately the study in \citet{Schneider_Cabrera_2006}  was restricted to the case of binary systems
composed of stars with equal masses, while a multitude of three-body configurations can be imagined.
In particular, we may wonder about the effect of a giant satellite orbiting an
already detected Jupiter-like extra-solar planet.
Moreover, the conclusions in \citet{Schneider_Cabrera_2006} were derived in the
framework of the restricted three-body problem and
we may also question the validity of this approximation.
Additionally, we found that their work contains several inaccurate conclusions.

Our goal is to correctly model the effect of a binary system on a nearby star
without making any assumptions on the masses of this triple system. 
In Sect.\,2 we briefly review the wobble method for detecting planets.
In Sect.\,3 we derive simple expressions for the radial velocity of a star
under the presence of a nearby binary system and for the orbital parameters of a
planet mimicked by this effect.
In Sect.\,4 we give some examples of binary systems and present numerical
simulations to test the theoretical results.
Finally, the last section is devoted to a discussion of the results.

\section{The wobble method}

\begin{figure}[h]
  \centering
    \includegraphics[width=6.5cm]{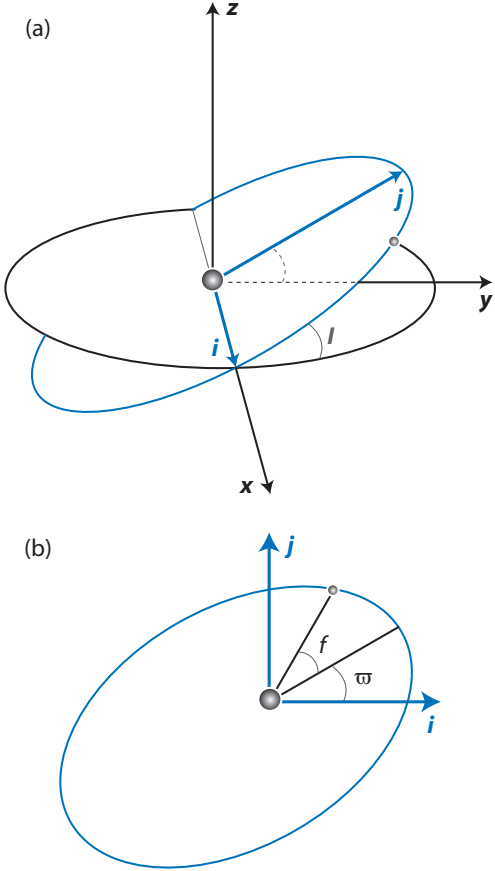}
  \caption{Reference frames. The plane defined by $(\vec{\hat x}, \vec{\hat y})$
  is the plane of the sky and the $ \vec{\hat z} $ axis is along the line of
  sight. The plane $(\vec{\hat \imath}, \vec{\hat \jmath})$ is the orbital
  plane, $ \varpi $ the argument of the perihelium and $ f $ the true anomaly.
  For simplicity we chose the $ \vec{\hat x} $ axis along the line of nodes ($
  \vec{\hat \imath} \equiv \vec{\hat x} $). \label{fig0}}
\end{figure}

We quickly review the wobble method for detecting extra-solar planets
\citep[see for instance][]{Hilditch_2001}. 
Assume that we are observing a star with mass $M_{\star}$, which has a planet
companion with mass $M_{p}$. A possible observation reference frame is depicted
in Fig.\ref{fig0}a where we have the $ \vec{\hat x} $ and $ \vec{\hat z} $ axes
along the line of nodes and the line of sight, respectively. 
The orbital plane $(\vec{\hat \imath}, \vec{\hat \jmath})$ is depicted in
Fig.\ref{fig0}b, where $ \vec{\hat \imath} \equiv \vec{\hat x} $.
In this last frame the position of the star/planet can be written as
\begin{equation}
\vec{r} = X \, \vec{\hat \imath} + Y \, \vec{\hat \jmath}
= r \cos (\varpi + f) \, \vec{\hat \imath} + r \sin (\varpi + f) \, \vec{\hat \jmath} \ ,
\end{equation}
where $r$ is the orbital radius, $f$ the true anomaly, and $\varpi$ the
longitude of the perihelium.

Since $ \vec{\hat \imath} \equiv \vec{\hat x} $, the $ \vec{\hat \jmath} $ axis
is simply obtained with a rotation of the $ \vec{\hat y} $ axis by an angle $I$,
the inclination of the orbital plane measured with respect to the plane
($\vec{\hat x},\vec{\hat y}$).
Thus, we may write $ z = Y \sin I $, where $z$ is the position along
the $ \vec{\hat z} $ axis. 
The velocity along the line of sight is then
\begin{eqnarray}
\dot{z} &=& \dot{Y} \sin{I}  \nonumber \\ 
        &=& (\dot{r}\sin(\varpi+f)+r\dot{f}\cos(\varpi+f)) \sin{I}  \\ 
        &=& K_z \left[ \cos(\varpi + f) + e \cos \varpi \right]  \nonumber \ ,
\label{vradial}
\end{eqnarray}
where $e$ is the orbital eccentricity and $ K_z = n \, a_p \sin I / \sqrt{1-e^2}
$, with $ n $ the mean motion, and $ a_p $ the semi-major axis of the orbit.

The radial velocity of the star is the projection of the velocity
with respect to the center of mass (CM) along the line of sight.
In the particular case of a circular orbit, we have $e=0$ and $ \dot f = n $,
thus 
\begin{equation}
\label{vradial0}
V_{r} =  \frac{M_p}{M_{\star}+M_{p}} \, \dot z = K\cos(\varpi + n \, t) \ ,
\end{equation}
where $ t $ is the time and the amplitude 
\begin{equation}
K=\frac{M_{p}}{M_{\star}+M_{p}} \, n \, a_p \sin{I} \ .
\end{equation}

If we analyze the Fourier spectrum of the star's radial velocity we will
see that this will exhibit a peak at the frequency $n$ and the amplitude $ K $
provides a lower limit for the planet's mass. 
This method of detection of extra-solar planets is also known as the wobble
method (since the planet causes the star to orbit or wobble about their common
CM). 

Notice also that when more than one companion to the star is present (another planet 
or star), the total radial velocity will be given by a sum of the
individual contribution from each body, $ V_r = \sum_k M_k \dot z_k / M_{Tot} $,
where $ M_{Tot} = M_\star + \sum_k M_k $ is the total mass of the system.
The amplitude due to a specific companion is then:
\begin{equation}
\label{kappa0}
K_p = \frac{M_{p}}{ M_\star + \sum_k M_k} \, n \, a_p \sin{I} \ .
\end{equation}


\section{The model and its predictions}

\label{modelpred}

\begin{figure}[h]
  \centering
    \includegraphics[width=8.5cm]{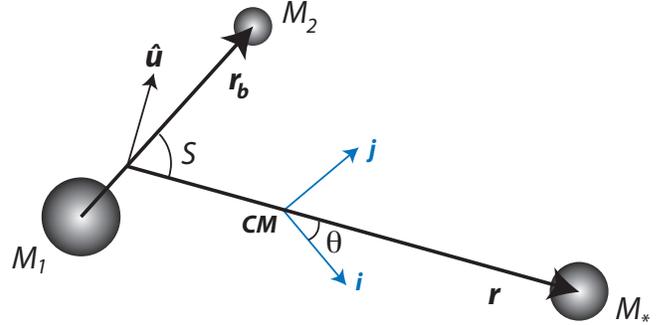}
  \caption{Jacobi coordinates ($\vec{r}_b, \vec{r}$). The system is composed by
  a star with mass $M_{\star}$ and a binary with masses $M_{1}$ and $M_{2}$. We
  assume co-planar motion and the frame $(\vec{\hat \imath}, \vec{\hat \jmath})$
  is the same from Fig.\ref{fig0}. $ \vec{\hat u} $ is a unit vector normal to
  $\vec{r}$, $S=\angle(\vec{r},\vec{r_b})$ and $\theta=\angle(\vec{\hat
  \imath},\vec{r})$. \label{fig1}}
\end{figure}

We will work in the framework of the planar general three body problem: we assume that
a star with mass $M_{\star}$ moves in the same orbital plane as the binary system composed of
two bodies with masses $M_{1}$ and $M_{2}$ (Fig.~\ref{fig1}).

\subsection{Equations of motion}

In order to obtain the equations of motion we will use Jacobi canonical
coordinates, which are, respectively, the inter-binary distance, $\vec{r_b}$, and
the distance, $ \vec{r} $, from the star $ M_{\star} $ to the binary's CM (see
Fig.~\ref{fig1}). 
We will additionaly assume that $|\vec{r_b}|  \ll  |\vec{r}|$. 
Thus, to the second order in $ \rho = |\vec{r_b}| / |\vec{r}| \ll 1$, the
Hamiltonian of the system can be expressed as
\citep[e.g.][]{Henrard_1979,Murray_Dermott_1999}:
\begin{eqnarray}
\label{hamiltonian}
 H & = &   \frac{1}{2} \frac{| \vec{p}_b |^2}{\mu} 
         + \frac{1}{2} \frac{| \vec{p} |^2}{\mu_\star} 
         - G \frac{M_{1} M_{2}}{r_b} 
         - G \frac{M_{} M_{\star}}{r} \\
   &   & - G \frac{M_{1} M_{2} M_{\star}}{r M_{}} \rho^2 
             \frac{1}{2} \left( 3 \cos^2 S - 1 \right)  \nonumber \ ,
\end{eqnarray}
where $ M_{} = M_1 + M_2 $ is the total mass of the binary,
$S=\angle(\vec{r},\vec{r_b})$,
\begin{equation}
\label{mus}
\mu = \frac{M_{1} M_{2}}{M_{}} \quad \mathrm{and} \quad \mu_\star = \frac{M_{}
M_{\star}}{M_{} + M_{\star}} 
\end{equation}
are the reduced masses and $ \vec{p}_b = \mu \, \dot{\vec{r}}_b $ and $ \vec{p} =
\mu_\star \, \dot{\vec{r}} $ the conjugate momenta of $ \vec{r}_b $ and $ \vec{r} $,
respectively.
In this framework, since the Jacobi coordinates are canonical, the equations of
motion are easily obtained as 
\begin{equation}
\label{motioneqham}
\dot{\vec{p}}_b = - \frac{\partial H}{\partial \vec{r}_b} \quad \mathrm{and}
\quad \dot{\vec{p}} = - \frac{\partial H}{\partial \vec{r}} \ .
\end{equation}

\subsection{Motion of the binary}

From Eq.(\ref{motioneqham}) the equation of motion for $\vec{r}_{b}$ is
\begin{eqnarray}
\label{jacobi1}
\ddot{\vec{r}}_b=-G\frac{M_{}}{r_{b}^3} \vec{r_b}+
	      G\frac{M_{\star}}{r^3}(3\rho \cos{S} \vec{r}-\vec{r_b}) \ .
\end{eqnarray}
We can neglect the second term in Eq.(\ref{jacobi1}) if $ \rho^3 \ll M_{} / M_{\star} $, which is always
true if we  choose a ``stable binary'', i.e.\ if $M_1 \ge M_2$, such that
\begin{equation}
\label{stablebinary}
\rho = \alpha \left( \frac{M_{1}}{3 M_{\star}} \right)^{1/3} \ ,
\end{equation}
where $\alpha \la 0.5$ \citep{Markellos_Roy_1981}.   
Therefore, the 0th order solution for $\vec{r_b}$ is a Keplerian ellipse with
constant semi-major axis $a_b$ and frequency 
\begin{equation}
\label{freq_om}
\omega=\sqrt{\frac{G M_{}}{a_{b}^3}} \ .
\end{equation} 

\subsection{Motion of the star}

From Eq.(\ref{motioneqham}) the equation of motion for $\vec{r}$ is
\begin{eqnarray}
\label{jacobi2}
\ddot{\vec{r}}& = & -G\frac{ (M_{} +M_{\star})}{r^3} \times  \\ 
              & & \left[ \left( 1+\frac{\mu}{M_{}}
\frac{\rho^2}{2} (-3+15 \cos^2{S}) \right) \vec{r} 
               -  \frac{\mu}{M_{}} 3\rho \cos{S} \vec{r_b} \right] \ . \nonumber
\end{eqnarray}
We can neglect the 2nd and 3rd terms in Eq.(\ref{jacobi2}) if 
\begin{equation}
\rho^2 \frac{\mu}{M_{}} \ll 1 \ .
\end{equation}
Since $\mu / M_{} \le 1/4 $ and from Eq.(\ref{stablebinary}) $\rho^2
\ll 1$, this is always true.
Therefore, the 0th order solution for $\vec{r}$ is also a Keplerian ellipse with
constant semi-major axis $a$ and frequency 
\begin{equation}
\label{freq_bigom}
\Omega=\sqrt{\frac{G (M_{} + M_{\star})}{a^3}} \ ,
\end{equation} 
i.e., the motion of $M_{\star}$ is to a first approximation a Keplerian orbit about a mass $M_{}$ located at the 
binary's CM.
Furthermore, from Eqs.~(\ref{freq_om}) and (\ref{freq_bigom})
we see that in general $\Omega \ll \omega$  since $ a_b \ll a $ 
(this is true unless $M_{\star}$ becomes too large).

\subsection{Stellar wobble}

Now, we can obtain an approximation for the relative motion of $M_{\star}$ by
replacing the 0th order approximations for $\vec{r_b}$ and $\vec{r}$  in
Eq.(\ref{jacobi2}). 
We assume that these 0th order solutions are circular orbits, thus  
\begin{eqnarray}
\vec{r} &=& a \, \vec{\hat r} \ , \\
\vec{r_b} &=& a_b\cos{S} \vec{\hat r} + a_b \sin{S} \vec{\hat u} \ , \\
 S &=& S_0 + (\omega-\Omega) t  \label{dotS} \ ,
\end{eqnarray}
where $\vec{\hat r}$ is the versor of $\vec{r}$, $\vec{\hat u}$ is the unit
vector orthogonal to $\vec{\hat r}$ and $S_0$ is an initial phase (Fig.\ref{fig1}).

In order to compute the radial velocity, we are interested in obtaining the
equation of motion for $M_{\star}$ in barycentric coordinates. 
The distance of the star to the CM is
\begin{equation}
\vec{r_{\star}} = \frac{M_{}}{M_{}+M_{\star}} \, \vec{r}  \ .
\end{equation}
We hence have
\begin{eqnarray}
\label{eqmotion1}
\ddot{\vec{r}}_{\star}& = & -G \frac{M_{}}{a^2} 
              \left[ \left( 1+\frac{\mu}{M_{}}
             \rho^2 (\frac{3}{4}+\frac{9}{4} \cos 2 S) \right) \vec{\hat r} \right.  \\
              & & \quad \quad \quad \quad - \left. \frac{\mu}{M_{}} \rho^2 \frac{3}{2}
	      \sin 2 S  \vec{\hat u} \right] \nonumber \ ,
\end{eqnarray}
with  $\rho=a_b/a  \ll 1$. 
Notice that if we chose $M_{\star}=0$ and $M_{1}=M_{2}$ (binary with equal
masses) in Eq.(\ref{eqmotion1}) we recover the same equation of motion obtained
by \citet{Schneider_Cabrera_2006}. 

In Appendix A we determine the solution to Eq.(\ref{eqmotion1}) in the inertial
frame $(\vec{\hat \imath}, \vec{\hat \jmath})$ with
coordinates $\vec{r_{\star}}=(X,Y)$, that is:
\begin{eqnarray}
\label{solution10}
\left(\begin{array}{c}
X \\ Y
\end{array}\right)
= 
\left(\begin{array}{c}
a_{\star} \cos \theta \\ a_{\star} \sin \theta  
\end{array}\right)
+ 
\left(\begin{array}{cc}
\sin \theta & \cos \theta \\ -\cos \theta & \sin \theta
\end{array}\right)
\left(\begin{array}{c}
3 \, \delta \sin 2 S \\  4.5 \, \delta \cos 2 S  
\end{array}\right) \ ,
\end{eqnarray}
with $a_{\star} \approx a M/(M+M_{\star})$, $ \theta = \theta_0 + \Omega \, t $ and
\begin{equation}
\delta=\frac{\mu}{8 M_{}} \left( \frac{a_b}{a} \right)^4 a_b \ .
\end{equation}
This solution has two components, the first one
caused by the motion of the star around the binary's CM and the second one is the
stellar wobble caused by the the movement of the two bodies of the binary around
its own CM.

\begin{figure}
  \centering
    \includegraphics[width=9.cm]{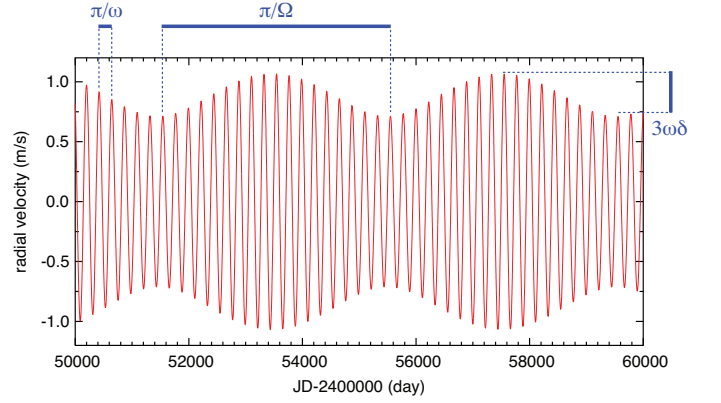}
  \caption{Radial velocity of a star due to the binary wobble
  around its own CM (Eq.\ref{doty4}). The orbital parameters are from
  Ex.\#3, Tab.\ref{table1}. \label{figmodul}}
\end{figure}

According to our choice of  reference frame (Fig.\ref{fig0}),
the radial velocity of the star $M_{\star}$ is given by $ V_{r} = \dot{Y} \sin{I}$, 
that is, we can obtain it directly from the $Y$ coordinate.
From Eq.(\ref{solution10}) we have:
\begin{equation}
\label{mimic1}
Y = a_{\star} \sin \theta -3 \, \delta \sin 2S \cos \theta + 4.5 \, \delta \cos 2S \sin \theta \ ,    
\end{equation}
Because $ \dot S = \omega - \Omega $ and  $ \dot \theta = \Omega $,
and since we assume $ \Omega \ll \omega $, the velocity along $ \vec{\hat y}$ is then
\begin{equation}
\label{doty2}
\dot{Y} = a_{\star} \Omega \cos \theta - 6 \omega \delta \cos 2S \cos \theta 
- 9 \omega \delta \sin 2S \sin \theta \ .  
\end{equation}
The first term in Eq.(\ref{doty2}) is due to the slow motion of the star around the binary's CM.
The second and third terms are due to the binary wobble. These last two terms can be seen as a 
composition of two periodic signals, 
one fast with frequency $ 2 \omega - 2 \Omega $ and another much slower with frequency $ \Omega $.
The global effect of the binary wobble corresponds to a signal of period $ \pi /
(\omega - \Omega) $ with an amplitude modulation of $ 3 \omega \delta $ and
period $ \pi / \Omega $ (Fig.\ref{figmodul}).

\subsection{Mimicking a planet}

\label{mimicking}

\begin{table*}
\centering
\caption{Planets detected in close binary systems and the respective
parameters ($a_b$, $M_2$) of the stellar companion that can mimic them.}             
\label{tabobs}      
\begin{tabular}{l|c c c|c c|c c c}        
\hline \hline
system name & $M_*$ ($M_\odot$) & $ M_1 $ ($ M_\odot $) & $M_p$
($M_\mathrm{Jup}$) & $a$ (AU) & $a_p$ (AU) & $ M_2 $ ($ M_\odot $) & 
$a_b$ (AU) &   $\alpha$  \\
\hline 
\noalign{\smallskip}
{\small GJ}\,86\,A$^{1}$ & 0.79 & $\sim$0.5 & 4.01 & $\sim$18 & 0.11 & $3.4 \times 10^3$ & 2.85 & 0.27 \\
$\gamma$\,Cep\,A$^{2}$ & 1.40 & 0.41 & 1.60 & 20.2 & 2.04 & 3.81 & 4.68 & 0.50 \\
{\small HD}\,41004\,A$^{3}$ & 0.7 & 0.4 & 2.54 & 23 & 1.31 & 11.9 & 5.41 & 0.41 \\
{\small HD}\,41004\,B$^{3}$ & 0.4 & 0.7 & 18.37 & 23 & 0.018 & $4.9 \times 10^5$ & 3.06 & 0.16 \\
{\small HD}\,196885\,A$^{4}$ & 1.33 & $\sim$0.6 & 2.96 & $\sim$23 & 2.63 & 2.93 & 5.78 & 0.47 \\
{\small HD}\,19994\,A$^{5}$ & 1.34 & 0.3 & 1.68 & $\sim$100 & 1.42 & $6.2 \times 10^2$ & 17.4 & 0.41 \\
\hline
 \noalign{\smallskip}
\end{tabular}

{\footnotesize References: [1] \citet{Queloz_etal_2000, Lagrange_etal_2006}; [2]
\citet{Neuhaeuser_etal_2007}; 

[3] \citet{Zucker_etal_2004}; [4]\citet{Correia_etal_2008}. [5]
\citet{Mayor_etal_2004}. }
\end{table*}

Even though the radial velocity of a star perturbed by a binary system is
not given by a single periodic signal (Fig.\ref{figmodul}), we may ask under
what circumstances this signal can be misinterpreted as a planet orbiting the
star. 
We can rewrite Eq.(\ref{doty2}) in the same format as
Eq.(\ref{vradial0}), i.e.
\begin{equation}
\label{doty3}
V_r  =  K_0 \cos (\theta_0 + \Omega \, t) + V_r^b \ ,
\end{equation}
where $K_0 = a_{\star} \Omega \sin I$ and
\begin{equation}
\label{doty4}
V_r^b  = K_1 \cos (\varpi_1 + n_1 t) + K_2 \cos (\varpi_2 + n_2 t) \ ,
\end{equation}
with
\begin{equation}
\label{kappa1}
K_1 = - \frac{15}{2} \omega \delta \sin I \quad \mathrm{and} \quad K_2 =
\frac{3}{2} \omega \delta \sin I \ ,
\end{equation}
\begin{equation}
\label{freqfond}
n_1 = 2 \omega - 3 \Omega \quad \mathrm{and} \quad 
n_2 = 2 \omega - \Omega \ ,
\end{equation}
and
\begin{equation}
\varpi_1 = 2 S_0 - \theta_0 \quad \mathrm{and} \quad 
\varpi_2 = 2 S_0 + \theta_0 \ .
\end{equation}

The 1st term in Eq.(\ref{doty3}) represents the slow two-body motion around a body
with mass $M$ located at the binary's CM. We recall that we may not know about the 
presence of the binary (since one or even both components may be unresolved) and this 
term allows us to identify only a single component.

On the other hand, $ V_r^b $ is the radial velocity due to the binary wobble
around its own CM and can be identified as two planets on circular orbits around the star.  
Hence, contrarily to \citet{Schneider_Cabrera_2006}, we conclude that the
stellar wobble caused by the presence of the binary mimics not one, but two
planets, although with very close orbital periods. 
As the two frequencies $n_1$ and $n_2$ are very close, in some situations the two
signals can be confounded and the observer can erroneously believe to have found
a planetary companion to the star.

This will be the case if our instrument is at the limit of its resolution: 
since $ | K_1 | = 5 \, | K_2 | $, the trace of the second ``planet'' is harder to detect.
Therefore, if our spectral resolution is close to the detection limit of the
stellar wobble, we can notice the presence of the first ``planet'', but we will
miss the second one.

By applying Kepler's 3rd law, we obtain for the semi-major axis of this fake
planet,
\begin{equation}
\label{aplanet}
a_{p} = \frac{(G M_{\star})^{1/3}}{n_1^{2/3}}  \approx \left(
\frac{M_\star}{4 M} \right)^{1/3} a_b \ ,
\end{equation}
and for the planet's mass, we replace $K_1$ from Eq.(\ref{kappa1}) in Eq.(\ref{kappa0}):
\begin{equation}
\label{mplanet}
\frac{M_{p}}{M + M_{\star}} = \frac{15}{2} \frac{\delta}{a_{p}}
\frac{\omega}{n_1}  \approx \frac{15}{32}
\frac{\mu}{M} \left( \frac{a_b}{a} \right)^4 \frac{a_b}{a_p} \ .
\end{equation}
Notice that when $M_1=M_2$ our estimate for the planet's mass is different from 
what was obtained by \citet{Schneider_Cabrera_2006}. Additionally, we also conclude that the planet evolves
in a circular orbit \citep[while $ e = 0.745 $ from Eq.4 of ][]{Schneider_Cabrera_2006}.


%
%

\subsection{Distinguishing between a binary and a planet}

The situation described above is very likely and we may
expect to interpret erroneously the signal of a binary as a planet.
However, the radial velocity wobble produced by a binary (Eq.\ref{doty4}),
is different from the radial velocity produced by a single planet
(Eq.\ref{vradial0}) and therefore we may be able to distinguish between the two
situations. 

Indeed, if we acquire data for long enough time (so that we can constrain well
$\Omega $) and if the precision of the instrument is at the limit of detection of
the amplitude $ K_2 $, the observer will be able to notice the presence of the
two periodic signals $ n_1 $ and $ n_2 $ (Eq.\ref{freqfond}) and therefore
realize that the star is undergoing the perturbation from a binary and not from
a planet.

Since the radial velocity of the binary (Eq.\ref{doty4}) is the same as the
signal produced by two planets in circular orbits, the observer can also believe
that two planets have been found (instead of the binary).
However, as long as $ \Omega \ll \omega $, the two orbital frequencies $ n_1 $
and $ n_2 $ are very close and probably the orbits of the two planets cannot be
stable.

For higher $ \Omega $ values, the system may become stable (although the
approximation $ \Omega \ll \omega $ may not be valid) and the hypothesis of
existence of two planets cannot be excluded.
Nevertheless, in the case of a binary, the frequencies ($ n_1 $, $ n_2 $) and
the amplitudes ($ K_1 $, $ K_2 $)  are not independent (Eqs.\ref{freqfond} and
\ref{kappa1}) and the following relations must be verified: 
\begin{equation}
n_2 - n_1 = 2 \Omega \quad \mathrm{and} \quad | K_1 | = 5 \, | K_2 | \ .
\end{equation}

\subsection{What kind of binary can mimic a planet?}

In Sect.\,\ref{mimicking} we saw that under some particular conditions, the
variations in the star's radial velocity resulting from the binary wobble can be
misinterpreted as a planet.  
Thus, if we detect a planet and we do not know about the presence of the binary
system (e.g. if $M_2$ is unresolved) we can ask the following question: 
what are the parameters of an hypothetical binary system that can mimic this planet? 
More precisely, if we know  $M_{\star}$, $M_1$, $a$, $M_p$ and $a_p$, 
what are the values of $M_2$ and $a_b$? 
These can be obtained by inverting Eqs.(\ref{aplanet}) and (\ref{mplanet}):
\begin{eqnarray}
\label{bplanet}
a_b &=& \frac{(G M)^{1/3}}{(\frac{1}{2} n_1 + \frac{3}{2} \Omega)^{2/3}} \approx \left(\frac{4
M_{}}{M_{\star}}\right)^{1/3} a_p \ , \\ 
\label{nplanet}
\mu \left( 1 + \frac{M_\star}{M} \right) &=& \frac{16}{15} \frac{n_1}{\omega}
\frac{a_p}{a_b} \left(\frac{a}{a_b} \right)^4 M_p  \\
&\approx& 0.212 \, \left( \frac{M_{\star}}{M} \right)^{5/3} \left(\frac{a}{a_p}
\right)^4 M_p \nonumber \ .   
\end{eqnarray}

In Table\,\ref{tabobs} we list all the currently known planets in close binary
systems ($ a < 100$~AU) and estimate the corresponding companions' parameters ($
M_2 $ and $ a_b $)
that can mimic that planet.
Three of the cases  provide unrealistic binary systems with very high values of $M_2$.
The remaining three casesare are acceptable ($ \gamma$\,Cep\,A,
{\small HD}\,41004\,A, and {\small HD}\,196885\,A) but these binary systems are likely to be unstable
as $\alpha$ is near the stability limit (see Eq.\ref{stablebinary}). On the other hand, 
the values of $M_2$ in these three cases are in the mass range of either black-holes or very bright stars (which should be detectable). 
Therefore, it is unlikely that these planets can be the result of a binary wobble.


\section{Examples}

In Sect.\,\ref{modelpred} we obtained the radial velocity wobble of a star that
is perturbed by a binary system.
We subsequently derived theoretical expressions that allow us
to determine the orbital parameters of a planet that can, in some circumstances,
be mimicked by those variations.
In this section we will present some concrete examples of binary systems 
together with the orbital parameters of the planets that they can mimic. We also obtain
the amplitude of the peaks in the radial velocity curve since these give an indication 
of the instrument precision needed to detect the planet and to distinguish between the planet
and binary.

\subsection{Theory versus numerical integrations}

In order to test the theoretical predictions of our model (Sect.\,\ref{modelpred}), we
performed some numerical simulations of three-body systems (Tab.\ref{table1}).
Here we present in detail a case (Ex.\#3, Tab.\ref{table1}) for which the current precision in the radial
velocity measurements ($ \sim$\,1~m/s) could lead to the erroneous announcement
of a planet.

The hypothetical system is formed by a triplet of main sequence stars with different
spectral types, G, K and M, and masses $ M_\star = 1.00\;M_\odot $, $ M_1 =
0.70\;M_\odot $ and $ M_2 = 0.35\;M_\odot $, respectively. 
The smaller K and M stars form a binary system on a circular orbit
with $ a_b = 1.1 $~AU. The G-star is in a wider circular orbit around the
close binary's CM, with $ a = 10 $~AU.
Note that the M star is much fainter than the K or G stars hence it represents
the unresolved component of the binary.
We also suppose that the system is co-planar and perpendicular to the line of
sight (that is, $ \sin I = 1 $).

We numerically integrated this stellar system and computed the radial
velocity of the G-star, the brightest body in the system.
In Table~\ref{vrdata} we list 50 simulated observational data points for a time
span of about 10.5~years (from Feb. 2004 to Sep. 2014).
We supposed that the data was acquired with a precision of $\sim$0.8~m/s, identical to
the {\small HARPS} spectrograph on the ESO 3.6-m telescope at La\,Silla (Chile),
currently the most precise instrument in operation since the beginning of 2004
\citep{Mayor_etal_2003}.

\begin{table}
\centering
\caption[]{Simulated radial-velocity data for a triple system of stars with $ M_\star = 1.00\;M_\odot $, $ M_1 =
0.70\;M_\odot $, $ M_2 = 0.35\;M_\odot $, $ a = 10.0$~AU and $ a_b = 1.1 $~AU.
The 50 points are generated for a period of about 10.5\,yr.} 
\label{vrdata} 
\begin{tabular}{c r c}        
\hline \hline
\textbf{Julian date} & \multicolumn{1}{c}{\textbf{RV}} & \textbf{RV error} \\ 
\textbf{[T - 2400000]} & \textbf{[\,km\,s$^{-1}$]}&\textbf{[\,km\,s$^{-1}$]} \\ 
\hline 
53063.   &    9.0534	&  0.0014\\   
53093.   &    9.2167	&  0.0007\\
53151.   &    9.5290	&  0.0011\\
53191.   &    9.7458	&  0.0010\\
53207.   &    9.8315	&  0.0009\\
53252.   &   10.0764	&  0.0009\\
53291.   &   10.2879	&  0.0015\\   
53435.   &   11.0649	&  0.0008\\   
53450.   &   11.1445	&  0.0014\\   
53520.   &   11.5204	&  0.0007\\    
53564.   &   11.7537	&  0.0012\\    
53618.   &   12.0344	&  0.0012\\   
53654.   &   12.2186	&  0.0009\\   
53809.   &   13.0006	&  0.0011\\   
53869.   &   13.2901	&  0.0019\\
53892.   &   13.3982	&  0.0009\\    
53934.   &   13.5962	&  0.0008\\
53986.   &   13.8352	&  0.0009\\
54029.   &   14.0281	&  0.0011\\
54161.   &   14.5877	&  0.0007\\
54206.   &   14.7691	&  0.0008\\
54306.   &   15.1462	&  0.0007\\
54349.   &   15.3000	&  0.0009\\
54360.   &   15.3383	&  0.0009\\
54544.   &   15.9165	&  0.0010\\
54570.   &   15.9898	&  0.0009\\
54651.   &   16.2000	&  0.0008\\
54954.   &   16.7522	&  0.0007\\
55000.   &   16.8023	&  0.0008\\
55079.   &   16.8711	&  0.0011\\
55102.   &   16.8882	&  0.0019\\
55309.   &   16.9182	&  0.0008\\
55321.   &   16.9145	&  0.0009\\
55363.   &   16.8964	&  0.0008\\
55415.   &   16.8640	&  0.0011\\
55447.   &   16.8376	&  0.0009\\
55753.   &   16.3807	&  0.0007\\
56030.   &   15.6524	&  0.0009\\
56075.   &   15.5070	&  0.0009\\
56084.   &   15.4762	&  0.0017\\
56132.   &   15.3149	&  0.0012\\
56186.   &   15.1237	&  0.0009\\
56380.   &   14.3606	&  0.0008\\
56460.   &   14.0186	&  0.0012\\
56482.   &   13.9205	&  0.0008\\
56755.   &   12.6247	&  0.0019\\
56774.   &   12.5278	&  0.0008\\
56830.   &   12.2441	&  0.0009\\
56856.   &   12.1124	&  0.0008\\
56920.   &   11.7809	&  0.0007\\
\hline
\end{tabular}
\end{table} 

From Eq.(\ref{freq_om}) and Eq.(\ref{freq_bigom}) we compute for this system:
\begin{equation}
\label{freqtheo}
\frac{2 \pi}{\omega} = 411.23 \;\mathrm{day} \quad \mathrm{and} \quad
\frac{2 \pi}{\Omega} = 8067.0 \;\mathrm{day} \ .
\end{equation}
The observational data do not fully cover the orbital period of the G-star
around the binary's CM, but we will be nevertheless able to estimate it,
since we are assuming a high level of precision in the data.

According to our model predictions (Sect.\,\ref{mimicking}), the radial
velocity of the G-star can be decomposed in three sinusoidal terms
(Eq.\ref{doty3}). 
One of large amplitude, $ K_0 = 6.907 $\,km/s, but varying slowly with period $ 2 \pi / \Omega $, 
and the other two terms varying much faster with periods (Eq.\ref{freqfond})
\begin{equation}
\label{freqsnum}
\frac{2 \pi}{n_1} = 222.64 \;\mathrm{day} \quad \mathrm{and} \quad 
\frac{2 \pi}{n_2} = 210.99 \;\mathrm{day} \ ,
\end{equation}
and small amplitudes (Eq.\ref{kappa1})
\begin{equation}
\label{ampsnum}
| K_1 | = 0.89 \;\mathrm{m/s} \quad \mathrm{and} \quad | K_2 | = 0.18
\;\mathrm{m/s} \ . 
\end{equation}
Since the precision of the instrument is supposed to be around $0.8$~m/s, it
should be able to detect the signal from $K_1$, but not from $K_2$.

\begin{figure}
   \centering
    \includegraphics*[height=8.5cm,angle=270]{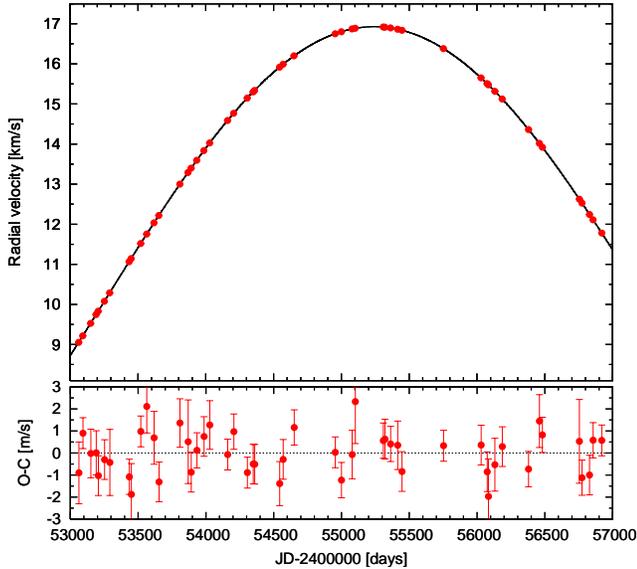}
  \caption{Radial velocities for the G-star superimposed on a Keplerian orbital
  solution due to the close binary CM. The remaining residuals are slightly
  above the precision of the instrument and the observer can suspect of the
  existence of second companion in the system. 
    \label{F1} }
\end{figure}

With the 50 data points listed in Table~\ref{vrdata} we will now apply the
traditional techniques used to search for planets.
The orbit of the binary CM introduces large amplitude variations in the radial
velocity of the G-star 
and at first glance we can only detect these variations.
Using the iterative Levenberg-Marquardt method \citep{Press_etal_1992},
we then first attempt to fit the complete set of radial velocities
with a single orbiting companion in a Keplerian orbit.
This fit yields a stellar companion with $P = 8010 $\,day, 
$e = 0.002 $ and a minimum mass of $ 1.044\;M_\odot$ (Fig.\ref{F1}).
These parameters correspond very well to the orbit of the G-star around the
binary's CM.
The mass determined by the fit is the total mass of the close binary system, 
$ M = M_1 + M_2 = 1.05 \; M_\odot $.

\begin{figure}
    \includegraphics*[height=8.5cm,angle=270]{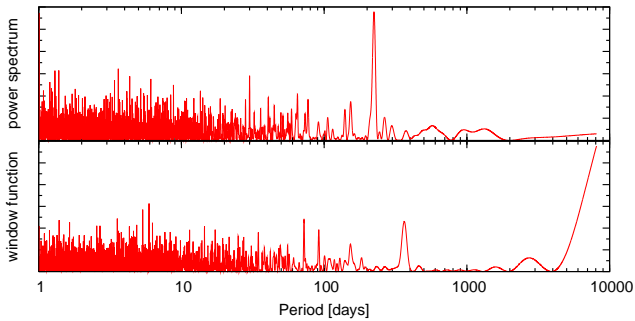} 
  \caption{Frequency analysis (top) and periodogram of the dates (bottom) for
  residual radial velocities when the contributions from the binary CM is
  subtracted. An important peak is detected at $P = 223$\,day, which can be
  interpreted as a second body in the system.
  \label{F3}}   
\end{figure}

In Figure~\ref{F1} we plot the radial velocity of the star superimposed on the
binary CM orbit and also the velocity residuals after subtracting these long
term variations.
The precision of the instrument is $\sim$1~m/s, but the residuals, although very
small, show amplitude variations that can reach almost $\pm $3~m/s.
A trained observer can then suspect for the existence of something else in the
data.
To check it in a simple way we perform a frequency analysis of the residuals
(Fig.\ref{F3}).
We can then see an important peak around 223~day, clearly indicating
that something else is hidden in the data.
We also plot a periodogram of the data, and since there is no peak at the same
frequency (there is only a small one around 364~day) we conclude that the signal
is true.

We then fit the radial velocities (Tab.\ref{vrdata}) using a model with two
Keplerian orbits. 
It yields for the planetary companion $P_1 = 222.6$~day, $e = 0.049$ and a minimum mass
of $20.6\,M_\oplus$, while for the binary's CM we get approximately the same
orbital parameters as before (Tab.\ref{T2}). 
Despite all the uncertainties in the orbital parameters, the use of a second
Keplerian orbit is a better approximation than the single Keplerian model, since the velocity residuals drop
to $rms$\,=\,$0.56$\,m/s, below the precision of the instrument (they were above
1~m/s before).

\begin{table}
\caption{Orbital parameters for two bodies orbiting the G-star, obtained with a
two-Keplerian fit to the observational data.}
\label{T2} 
\begin{center}
\begin{tabular}{l l c c} \hline \hline
\noalign{\smallskip}
{\bf Param.}  & {\bf [unit]} & {\bf planet} & {\bf binary CM} \\ \hline 
\noalign{\smallskip}
$V_0$        & [km/s]               & \multicolumn{2}{c}{$ 10.021 \pm 0.037 $}  \\  
$P$          & [day]                & $ 222.61 \pm 1.22  $ & $ 8018.0 \pm 32.5  $ \\ 
$e$          &                      & $ 0.0486 \pm 0.1725$ & $ 0.0016 \pm 0.0002$ \\ 
$\omega$     & [deg]                & $  94.04 \pm 237.2 $ & $ 280.35 \pm 34.93 $ \\ 
$K$          & [m/s]                & $ 1.07 \pm 0.19	 $ & $ 6904.6 \pm 30.1  $ \\  
$T$          & [JD-2400000]         & $ 53839.0 \pm 146.9$ & $ 53471.9 \pm 778.0$ \\  \hline
\noalign{\smallskip}
$a_1 \sin i$ & [$10^{-3}$ AU]       & $ 0.022 $          & $ 5089  $ \\
$f (m)$      & [$10^{-3}$ M$_\odot$]& $ 2.8 \times 10^{-11}  $           & $ 273.5 $ \\
$m_2 \sin i$ & [M$_\oplus$]         & $ 20.6 $           & $ 3.5 \times 10^5 $ \\
$m_2 \sin i$ & [M$_\odot$]          & $ 6.2 \times 10^{-5} $        & $ 1.046 $ \\
$a$          & [AU]                 & $ 0.719 $           & $ 9.95 $ \\ \hline
\noalign{\smallskip}
$rms$        & [m/s]                & \multicolumn{2}{c}{0.56}   \\
$\sqrt{\chi^2}$     &                      & \multicolumn{2}{c}{0.51}   \\ \hline
\noalign{\smallskip}
\end{tabular}
\end{center}
{\footnotesize Errors are given by the standard deviation $ \sigma $. The orbital period of the
outer body is longer than the data acquired and thus we are unable to complete
constrain its orbit. The eccentricity and the perihelium of the inner planet are
also very badly constrained because we are at the limit of the instrument
resolution. Despite all the uncertainties, the adjusted parameters for the
planet agree with our model predictions (Sect.\,\ref{mimicking}).}
\end{table}

\begin{figure}
    \includegraphics*[height=8.5cm,angle=270]{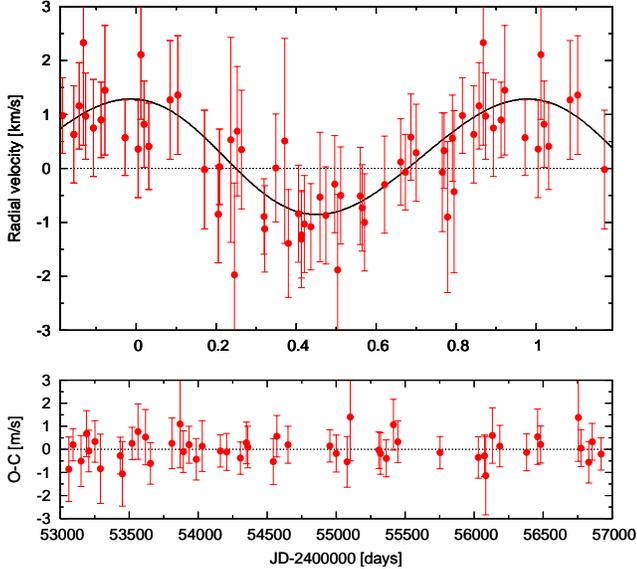} 
  \caption{Phase-folded residual radial velocities when the contribution of the
  binary CM is subtracted. Data is superimposed on a Keplerian solution 
    with $P = 222.6$~day and $ e \approx 0 $. The respective residuals as a function of
    Julian Date are displayed in the lower panel. We see that the remaining
    $O-C$ is below the precision of the instrument ($\sim$1~m/s) and therefore
    we do not expect to find any additional signal in the data. 
    \label{F2}}   
\end{figure}

In Figure~\ref{F2} we plot the phase-folded residual radial velocities when the
contribution of the binary CM is subtracted. 
The data is superimposed on the orbital solution of the planetary companion,
which really looks like a planet.
The remaining $O-C$ is below the precision of the instrument ($\sim$1~m/s) and
therefore we do not find any additional signal in the data.
This is confirmed when we perform a frequency analysis of these residuals: no
relevant peak is found, and there is no signal around $P_2 = 211 $~day, the period
for which we expected to see the $ n_2 $ frequency (Eq.\ref{freqsnum}).

\begin{table*}
\caption{Model predictions for different examples of binary systems around a star
with $ M_\star = M_\odot $.}             
\label{table1}      
\centering                          
\begin{tabular}{c|c c c c c|c c c |c c|c c}        
\hline\hline                 
   & \multicolumn{5}{c|}{Binary system} & \multicolumn{3}{c|}{Frequencies} & \multicolumn{2}{c|}{Amplitudes} & \multicolumn{2}{c}{Planet}  \\  
 Ex. & $ M_1$ & $M_2$ & $a$ & $a_b$ & $\alpha$ & $2\pi/\Omega$ &  $2\pi/n_1 $ & 
 $2\pi/n_2$ & $| K_1 |$ & $| K_2 |$ & $M_{p}$ & $a_{p}$ \\  
  & ($M_\odot$) & ($M_\odot$) & (AU) & (AU) & & (yr) &  (day) &  (day) & (m/s) & (m/s) & ($M_\oplus$) & (AU) \\  
\hline                        
 1  & 1.00 & 1.00 & 10.0 & 1.00 & 0.114 & 18.26 & 137.1 & 131.7 & 0.987 & 0.197 & 23.90 & 0.520  \\  
 2  & 1.00 & 1.00 & 10.0 & 1.50 & 0.172 & 18.26 & 265.6 & 246.0 & 4.081 & 0.816 & 123.1 & 0.809  \\  
 3  & 0.70 & 0.35 & 10.0 & 1.10 & 0.156 & 22.09 & 222.6 & 211.0 & 0.888 & 0.178 & 17.26 & 0.719  \\ 
 4  & 1.00 & 0.10 & 10.0 & 1.50 & 0.210 & 21.82 & 363.7 & 333.3 & 1.000 & 0.200 & 23.47 & 0.997  \\  
 5  & 1.00 & 0.01 & 10.0 & 1.50 & 0.215 & 22.30 & 380.6 & 348.1 & 0.114 & 0.023 &  2.59 & 1.028  \\  
 6  & 1.00 & $10^{-3}$ & 10.0 & 1.50 & 0.216 & 22.36 & 382.5 & 349.7 & 0.012 & 0.002 & 0.26 & 1.031  \\    
 7  & $10^{-3}$ & $10^{-3}$ & 1.00 & 0.01 & 0.114 & 1.00 & 4.23 & 4.13 & $10^{-5}$ & $10^{-6}$ & $10^{-4}$ & 0.051 \\  
 8  & $10^{-3}$ & $10^{-6}$ & 1.00 & 0.01 & 0.144 & 1.00 & 6.06 & 5.86 & $10^{-7}$ & $10^{-8}$ & $10^{-7}$ & 0.065 \\  
\hline      	  	     	 
\end{tabular}
\end{table*}

By comparing the numerical determination period of the planet, $ P_1 = 222.61 \pm
1.22 $~day (Tab.\ref{T2}) with its theoretical estimation $ P_1 = 222.64 $~day
(Eq.\ref{freqsnum}), we see that our prediction was correct.
Results for the amplitude are not so good, but still exact since the
numerical determination $ K_1 = 1.07 \pm 0.19 $~m/s (Tab.\ref{T2}) includes 
within the error bar the theoretical estimation $ K_1 = 0.89 $~m/s
(Eq.\ref{ampsnum}).
A better numerical determination could be achieved with identical instrumental
precision, but using a larger amount of data points.
On the other hand, the amplitude $ K_2 = 0.18 $~m/s (Eq.\ref{ampsnum}) is lower
than $ \sigma_{K_1} = 0.19 $~m/s (Tab.\ref{T2}), the error bar of $ K_1 $, so
there is no hope of determining it with the present experience.

From the numerical values of $ n_1 $ and $ K_1 $ we can infer the
orbital parameters of the planet, $ a_p = 0.719 \pm 0.003 $~AU and $ M_p = 20.6
\pm 3.7 \;M_\oplus$, while the theoretical values were respectively $ a_p =
0.719 $~AU (Eq.\ref{aplanet}) and $ M_p = 17.3 \;M_\oplus $ (Eq.\ref{mplanet}).
The mass is overestimated because the amplitude $ K_1 $ was so.

Now that we know that a binary can mimic a planet, whenever we suspect that 
a binary can be responsible for the observed stellar wobble, we should also
provide its parameters.
From the values of $ n_1 $ and $ \Omega $ (Tab.\ref{T2}) we can obtain
(Eq.\ref{freqfond})
\begin{equation}
\frac{2 \pi}{\omega} = 411.0 \pm 2.2 \;\mathrm{day} \quad \mathrm{and} \quad
\frac{2 \pi}{\Omega} = 8018 \pm 33 \;\mathrm{day} \ , 
\end{equation}
which are in very good agreement with the true values (Eq.\ref{freqtheo}).
The individual masses of the stars in the close binary can also be
resolved as well as the semi-major axis.
Directly from the fit we have $ a = 9.95 \pm 0.03$~AU and $ M = 1.046 \pm 0.005
\; M_\odot $ (Tab.\ref{T2}). 
Then, using Eqs.(\ref{bplanet}) and (\ref{nplanet}) we can respectively obtain
\begin{equation}
a_b = 1.098 \pm 0.006 \;\mathrm{AU} \quad \mathrm{and} \quad
\mu = 0.275 \pm 0.062 \; M_\odot \ . 
\end{equation}

By comparing the above determinations with the true parameters chosen for this
system (Ex.\#3, Tab.\ref{table1}) we verify that they match quite well.
We thus conclude that a two-Keplerian approach of the stellar radial velocity
is enough to determine the parameters of the triple system, i.e., we can keep
the current software that is used to fit planets to the stellar data and do
not need to develop a specific algorithm for binaries.
Of course this in only true if we keep a Keplerian approach, the use of a
N-body algorithm will give different results in the case of a planet or a
binary.

\subsection{Different kinds of binary systems}

Now that we have seen that our formulae are correct we will apply them to a wide
variety of binary systems' configurations.
In Table~\ref{table1} we see the predictions of the model for several binary
systems around a star with $M_{\star}=M_{\sun}$. We chose only stable binary
systems with $\alpha<<0.5$ (see Eq.\ref{stablebinary}).

The first and second examples are the cases studied in \citet{Schneider_Cabrera_2006}: the star is being perturbed 
by close equal masses' binaries composed of stars with $1\,M_{\sun}$. In Ex.\#1,  $K_1$ is at
the current resolution level ($\sim$1\,m/s) while $K_2$ is 5 times smaller. In this case we may be
lead to believe that the star has a planet companion with 1.4 Neptune masses. On
the other hand, in Ex.\#2 we should be able to detect both $K_1$ and $K_2$;
hence in this case we can conclude that the star's wobble is caused by a binary
and not a planet companion. 

The third example is the case discussed in the previous section. Here we have a triple system composed of stars
with different spectral types. As we saw previously, in this case we can detect
the peak with amplitude $K_1$, but not the one with amplitude $K_2$, and we are
lead to believe that the star has a planet companion  with about one Neptune mass. 

In the fourth example the star is being perturbed by a close hierarchical binary
composed of a star with $1\,M_{\sun}$ and another body with $0.1\,M_{\sun}$. We have $K_1$ at the current resolution level
while  $K_2$ is 5 times smaller. Again we are lead to believe that the star has a planet companion with 1.4 Neptune masses.

In the next two examples the binaries are composed of a star with $1\,M_{\sun}$
and a brown-dwarf with 
$10\,M_{J}$ (Ex.\#5) or a planet with $1\,M_{J}$ (Ex.\#6). 
In these cases both peaks are currently undetectable ($\sim$\,cm/s)
but could be within reach of future observation campaingns like CODEX
\citep{Pasquini_etal_2008}.

Finally, in the last two examples we have a $1\,M_{\sun}$ star and a binary composed of
two Jupiter-sized planets (Ex.\#7) and a Jupiter-sized planet and an
Earth-sized ``satellite'' (Ex.\#8).
In both cases, the inferred values for $M_p$ are too low for planets.



\section{Discussion and conclusion}

In this paper we studied a triple system composed of a star which we observe
and a binary system.  
Our goal was to decide if the effect of the binary system on the star's center
of mass motion could be mistaken as a planet companion to it.  
 
Previous work done by \citet{Schneider_Cabrera_2006} suggested that the
effect of a star-star system could mimic the presence of a planet companion to
the observed star. 
However, we found that this is not correct, since a more careful analysis of the
observational data should allow us to distinguish between the two effects.
Indeed, when there is a close binary system, the Fourier spectrum of the star's
radial velocity should exhibit not one but two peaks located at nearby
frequencies. This could also indicate the presence of two planets on close orbits,
but since these systems are most likely unstable we should be able to reject
this hypothesis.  

Nevertheless, we saw that it is sometimes possible to mistake (at least temporarily) the effect of a binary
system by a planet. This happens when the observations are near the resolution limit so that
we can identify the primary peak but not the secondary peak. 
However, even in this case our estimate for the fake planet's mass differs from that obtained by
\cite{Schneider_Cabrera_2006}. 

In this article we obtained tools for
distinguishing between the two effects: we  estimate the parameters of the binary system 
that could me mistaken by that planet and check if they are realistic.
We recommend that, in future discoveries of planets around stars in binary
systems, these formulas should be used in order to exclude the possibility of a
false alarm.

We also wanted to evaluate the effect of different types of binary systems on the star's CM motion.
In particular, we were interested in star-star, star-planet and planet-satellite systems.
We saw that the effect of a star-star system is only significant when this is close to
the observed star. For instance, a binary system of dwarf stars at 1~AU from each other and at
the distance 10~AU from the observed star causes a wobble on the latter 
which would be at the current resolution level.
We also saw that the effect of a star-planet system is currently undetectable although it is possibly
within reach of future observation campaigns.
Finally, we saw that the effect of a planet-satellite system is insignificant and should never be mistaken 
by a planet companion to the observed star as the implied mass would simply be
too low. 

Our conclusions were established for a triple system of bodies evolving in
co-planar and circular orbits. 
However, it is known that these kind of systems may present eccentric orbits and
different mutual inclinations.
The inclusion of these two effects may change some of the results presented
here.

\appendix

\section{}

We re-write the equation of motion of the star $M_{\star}$  (Eq.\ref{eqmotion1}) using Cartesian coordinates in the
$(\vec{\hat \imath}, \vec{\hat \jmath})$ frame (see Fig.~1) such that
\begin{eqnarray}
\vec{\hat r} &=& \cos \theta \, \vec{\hat \imath}+\sin \theta \, \vec{\hat \jmath} \ , \\
\vec{\hat u} &=& -\sin \theta \, \vec{\hat \imath}+\cos \theta \, \vec{\hat \jmath} \ ,
\end{eqnarray} 
with
\begin{equation}
\label{teta}
\theta = \theta_0 + \Omega \, t  \ ,
\end{equation}
where $\theta_0$ is an initial phase (Fig.\ref{fig1}).

Thus we obtain
 \begin{eqnarray}
      \label{ddotX}
      \ddot{X}  & = & -\frac{G M_{}}{a_{A}^2} \left[ 
      \left(1+\frac{3}{4} \rho^2 \frac{\mu}{M_{}} \right) \cos \theta  \right. \\
      \nonumber
       & + &  \left. \frac{9}{4} \rho^2 \frac{\mu}{M_{}} \cos 2 S \cos \theta 
       +  \frac{3}{2} \rho^2 \frac{\mu}{M_{}} \sin 2 S \sin \theta \right] \ , \\
      \label{ddotY}
      \ddot{Y}  & = & -\frac{G M_{}}{a_{A}^2} \left[
      \left(1+\frac{3}{4} \rho^2 \frac{\mu}{M_{}} \right) \sin \theta  \right. \\
      \nonumber
       & + & \left. \frac{9}{4} \rho^2 \frac{\mu}{M_{}} \cos 2 S \sin \theta 
       \nonumber
        -  \frac{3}{2} \rho^2 \frac{\mu}{M_{}} \sin 2 S \cos \theta \right] \ ,
   \end{eqnarray}
where $\rho=a_{b}/a \ll 1$.

In order to solve this set of differential equations we perform the transformation of coordinates
\begin{equation}
\left(\begin{array}{c}
\bar{X} \\ \bar{Y}
\end{array}\right)
=
\left(\begin{array}{cc}
\sin \theta & -\cos \theta \\ \cos \theta & \sin \theta 
\end{array}\right)
\left(\begin{array}{c}
X -a_{\star} \cos \theta  \\ Y - a_{\star} \sin \theta  
\end{array}\right) \ ,
\end{equation}
with 
\begin{equation}
\label{astar}
a_{\star} = \frac{M_{}}{M_{} + M_{\star}} \left( 1+\frac{3}{4} \rho^2 \frac{\mu}{M{}} \right) \, a  \ .
\end{equation}

The set of differential equations becomes 
\begin{eqnarray}
\ddot{\bar{X}}-2\Omega \dot{\bar{Y}} -\Omega^2 \bar{X} &=& -12\omega^2 \delta \sin 2 S \ , \\
\ddot{\bar{Y}}+2\Omega \dot{\bar{X}} -\Omega^2 \bar{Y} &=& -18\omega^2 \delta \cos 2 S \ ,
\end{eqnarray}
with
\begin{equation}
\delta=\frac{\mu}{8 M_{}} \left( \frac{a_b}{a} \right)^4 a \ .
\end{equation}

This admits the solution
\begin{eqnarray}
\bar{X} &=& \delta_{X} \sin 2 S \ , \\
\bar{Y} &=& \delta_{Y} \cos 2 S \ ,
\end{eqnarray}
where
\begin{eqnarray}
\label{eq1x}
-4(\omega-\Omega)^2 \delta_{X}+4\Omega(\omega-\Omega) \delta_{Y}-\Omega^2
\delta_{X} &=& -12\omega^2 \delta \ ,\\
\label{eq1y}
-4(\omega-\Omega)^2 \delta_{Y}+4\Omega(\omega-\Omega) \delta_{X}-\Omega^2
\delta_{Y} &=& -18\omega^2 \delta \ .
\end{eqnarray}

The solution to Eqs.~(\ref{eq1x}) and (\ref{eq1y}) is then
\begin{eqnarray}
\delta_{X} &=& 12\delta \frac{4 (1-\beta)^2+\beta^2+6\beta(1-\beta)}{(4(1-\beta)^2-\beta^2)^2} \ , \\
\delta_{Y} &=& 12\delta \frac{6 (1-\beta)^2+3\beta^2/2+4\beta(1-\beta)}{(4(1-\beta)^2-\beta^2)^2} \ .
\end{eqnarray}
with $\beta=\Omega/\omega \ll 1$.

To zero order in $\beta$,
\begin{eqnarray}
\delta_{X} &=& 3\delta \ , \\
\delta_{Y} &=& 4.5\delta \ ,
\end{eqnarray}
hence
\begin{eqnarray}
\label{solution1}
\left(\begin{array}{c}
X \\ Y
\end{array}\right)
= 
\left(\begin{array}{c}
a_{\star} \cos \theta \\ a_{\star} \sin \theta  
\end{array}\right)
+ 
\left(\begin{array}{cc}
\sin \theta & \cos \theta \\ -\cos \theta & \sin \theta
\end{array}\right)
\left(\begin{array}{c}
3 \, \delta \sin 2 S \\  4.5 \, \delta \cos 2 S  
\end{array}\right) \ .
\end{eqnarray}

Note that this does not match the solution obtained in \citet{Schneider_Cabrera_2006}:
\begin{itemize}
\item The 2nd term in Eq.(\ref{solution1}) is rotated by an angle $\theta$. 
\item The frequency of the 2nd term is $2(\omega-\Omega)\approx 2\omega$ as we assume $\Omega \ll \omega$.
\item The expressions for  $\Omega$ (Eq.\ref{freq_bigom}) and $a_{\star}$ (Eq.\ref{astar}) take into account the mass $M_{\star}$.
\end{itemize}

\begin{acknowledgements}
M.H.M.~Morais acknowledges financial support from FCT-Portugal (grant
SFRH/BPD/19155/2004).       
\end{acknowledgements}

\bibliographystyle{aa}
\bibliography{correia}
 
\end{document}